# White Paper on Enabling Global Image Data Sharing in the Life Sciences

Contributors

Peter Bajcsy (1) Sreenivas Bhattiprolu (2), Katy Börner (3), Beth A Cimini (4), Lucy Collinson (5), Jan Ellenberg  (6), Reto Fiolka (7), Maryellen Giger (8), Wojtek Goscinski (9), Matthew Hartley  (10), Nathan Hotaling (11), Rick Horwitz (12), Florian Jug (13), Anna Kreshuk (6), Emma Lundberg (14), Aastha Mathur (6), Kedar Narayan (15), Shuichi Onami (16), Anne L. Plant (1), Fred Prior (17), Jason Swedlow (18), Adam Taylor (19), and Antje Keppler (6)

Affiliations:
1. National Institute of Standards and Technology, Gaithersburg, USA.
2. ZEISS Microscopy Customer Center, Dublin, USA
3. Intelligent Systems Engineering, Indiana University, Bloomington, USA
4. Imaging Platform, Broad Institute, Cambridge, MA USA.
5. Electron Microscopy Science Technology Platform, Francis Crick Institute, 1 Midland Road, London, UK.
6. European Molecular Biology Laboratory, Heidelberg, Germany
7. Lyda Hill Department of Bioinformatics, UT Southwestern Medical Center, Dallas, Texas, USA.
8. Department of Radiology and Committee on Medical Physics, University of Chicago, Chicago, IL USA
9. National Imaging Facility, Brisbane, Australia
10.  European Molecular Biology Laboratory, European Bioinformatics Institute, Hinxton, Cambridge, UK
11.  National Center for Advancing Translational Science, National Institutes of Health, Rockville, USA
12.  Allen Institute for Cell Science, Seattle, USA
13.  Fondazione Human Technopole, Milan, Italy
14. Stanford University, California, US and SciLifeLab, KTH Royal Institute of Technology, Stockholm, Sweden
15.  Center for Molecular Microscopy, Center for Cancer Research, National Cancer Institute, National Institutes of Health, Bethesda USA. and Cancer Research Technology Program, Frederick National Laboratory for Cancer Research, Frederick USA.
16.  RIKEN Center for Biosystems Dynamics Research, Kobe, Japan
17.  Department of Biomedical Informatics, Department of Radiology, University of Arkansas for Medical Sciences, Little Rock, USA
18.  Divisions of Computational Biology and Molecular, Cell and Developmental Biology, University of Dundee, UK
19.  Sage Bionetworks, Seattle, WA, USA

**Abstract**

Coordinated collaboration is essential to realize the added value of and infrastructure requirements for global image data sharing in the life sciences. In this White Paper, we take a first step at presenting some of the most common use cases as well as critical/emerging use cases of (including the use of artificial intelligence for) biological and medical image data, which would benefit tremendously from better frameworks for sharing (including technical, resourcing, legal, and ethical aspects). In the second half of this paper, we paint an 'ideal world scenario' for how global image data sharing could work and benefit all life sciences and beyond. As this is still a long way off, we conclude by suggesting several concrete measures directed toward our institutions, existing imaging communities and data initiatives, and national funders, as well as publishers. Our vision is that within the next ten years, most researchers in the world will be able to make their datasets openly available and use quality image data of interest to them for their research and benefit. This paper is published in parallel with a companion White Paper entitled "*Harmonizing the Generation and Pre-publication Stewardship of FAIR Image Data*", which addresses challenges and opportunities related to producing well-documented and high-quality image data that is ready to be shared. The driving goal is to address remaining challenges and democratize access to everyday practices and tools for a spectrum of biomedical researchers, regardless of their expertise, access to resources, and geographical location.

## 1. Motivation for White Paper

Public, reference data are one of the most important foundational resources in the modern life and biomedical sciences. Over the last 40 years, the public release and availability of genomic and macromolecule structural databases accelerated discovery, sparked the development of wholly new fields of science, spawned multi-billion dollar industries, and led to a revolution in drug development and disease treatments. The major next step is the development of public reference biological and medical imaging data repositories, which promises at least an equal, if not larger impact on discovery and society. The motivation driving this White Paper is the timely and critical need for infrastructure supporting image acquisition, management, and analysis, that is not only coordinated within the US but also across the world.

Australia, Japan, and the European Union have established nationally funded, coordinated research infrastructures (RIs) for life science and biomedical imaging (collectively "BioImaging"). These BioImaging RIs are the output of concerted long-term strategic planning efforts from academic, funding, political as well as industrial partners, i.e., the key stakeholders in life sciences and biomedical research. In 2023, national and transnational BioImaging RIs are in operation and deliver technology and community support for quality BioImage data acquisition, management, analysis, and publication for their stakeholders.

As just one example, Euro-BioImaging ERIC, the European BioImaging RI[1], provides imaging technology platforms and data services to the European life and biomedical sciences community. One of the key provisions of Euro-BioImaging Image Data Services is the commitment to open sharing of data following the FAIR (Findable, Accessible, Interoperable, Reuseable) principles (Wilkinson et al., 2016). The work on Euro-BioImaging initiated in 2008 and the infrastructure finally became operational in 2019, requiring over a decade to coordinate such diverse, dynamic activities across a large number of countries and scientific communities. Nonetheless, the result is a coordinated effort where image data acquisition at advanced technology facilities across the continent are linked to public data resources, i.e., the BioImage Archive and the Image Data Resource, constructed and operated with significant investments from the UK and EU (Table 1). Publicly funded projects and community initiatives are also actively building and supporting image data formats[2] and applications for managing, analysing, and sharing data using the most advanced technologies, including AI[3]. Collaboration between academic and commercial organizations is maturing and provides a powerful ecosystem for development and eventual scaling of new technologies and products. Whereas Euro-BioImaging's Image Data Services are well-developed, they are also evolving and maturing to match the innovation of the rapidly developing field of BioImaging. Additionally, these services are being incorporated into the landscape of FAIR and interoperable data and services across Europe, by utilising, and hence constituting, a crucial component of the European Open Science Cloud (EOSC[4]) - a shared resource that aims to federate valuable digital resources and outputs across disciplinary and national boundaries.

---

[1] www.eurobioimaging.eu

Looking at the landscape in North America, there are early examples of engagement of the biological and medical imaging community e.g. BioImaging North America (BINA), Association of Biomolecular Resource Facilities (ABRF), and others (Table 3). However, at present, there is no route towards a common data infrastructure for coordinated image data management and analysis resources for the life sciences and bioimaging communities in the US or Canada. Because of this a trans-continental commitment to a common infrastructure for BioImaging is not on the horizon. Indeed, NIH alone has created several siloed biological and medical imaging data resources through their Common Fund and extramural programs. In fact, recently, the NIBIB of the NIH created MIDRC (the Medical Imaging and Data Resource Center), which is a multi-institutional collaborative initiative driven by the medical imaging community that was initiated in late summer 2020 to help combat the global COVID-19 health emergency. MIDRC is generalizable, scalable, and interoperable for multiple imaging use cases. Its aim is to foster machine learning innovation through data sharing for rapid and flexible collection, analysis, and dissemination of imaging and associated clinical data by providing researchers with unparalleled resources. However, once established resources like these become hard to reconcile, coordinate, and harmonize with other existing or future data resources in, as well as outside, the USA. Therefore, opportunities for discoveries through data integration, one of the key outputs from genomic and structural data resources, are missed and the return on large research investments reduced. Furthermore, there is a risk that contributions from North American science to the emerging technologies enabled by generative AI are reduced or possibly even missed all together.

To move from isolated efforts to a coordinated design and strategy for a sustainable image data resource, the academic and industrial community of imaging scientists need to join forces and engage with the national stakeholders and funding agencies to develop a common understanding of the scientific community's requirements and the desired outcomes of the regional and national funding institutions. Such a coordinated collaboration is similar to the agreement by the medical imaging industry to use the DICOM format (Bidgood et al., 1997), making medical images obtained from around the world accessible to all. In addition, the lesson of the European (and indeed the Australian, the Japanese, and several other countries) experience is that a connected, coordinated effort is required to build the image data research infrastructures that provide services to the wider community. Assuming that individual groups will provide sustainable infrastructure for the whole scientific community using standardized responsive funding mechanisms seems naïve and has been shown to be inadequate many times.

---

[2] https://ngff.openmicroscopy.org/
[3] https://ai4life.eurobioimaging.eu/
[4] https://www.eosc.eu/

Here, we suggest that imaging scientists, funders, publishers, technology developers, and vendors in North America come together to create this infrastructure and the training needed to support it. In our experience the establishment of several proof-of-concept (PoC) projects provides a rapid and powerful way to test different ideas for technology development, delivery, access, training, etc. and build the foundation for the future development of an accessible and useful data infrastructure. We include several specific use cases that would be candidates for these PoCs. It is notable that the priorities of each use case differ in detail, as the scope of the scientific applications that drive requirements in biological and medical fields varies. This demonstrates the challenge and opportunity of building a common BioImaging Data Infrastructure.

The European and other international communities of imaging scientists stand ready to support, advise, exchange ideas, and even to contribute to these efforts as the North American community initiates this essential journey for a common foundational BioImaging Data Infrastructure.

## 2. Background

Image data is the fastest growing data resource in the life sciences. The data are complex and deep in information, and the scientific community is only at the beginning to tap into this exponentially growing resource. Sharing FAIR quality-managed image data (Kemmer et al., 2023) widely supports open science. It increases reproducibility and transparency, facilitates collaboration, allows combining and analyzing datasets in new ways, inspires novel research questions and approaches, and most importantly, leads to innovation and new discoveries.

Many of the most exciting opportunities for deriving value from the reuse of biological and medical imaging data at scale rely on the ability to curate and aggregate those data so that they can be addressed together as a coherent whole. This aggregation enables very large scale AI/ML model training, as well as the creation of cross-modality and cross-domain benchmarks and reference datasets. It also opens the path for widely distributed computing, flexible movement of compute power to data or vice-versa, and mirroring of highly-accessed datasets to enable fast regional access, among other opportunities.

These gains are not possible when data are siloed and fragmented in many different places, often with no public access or consistent organization. Access to data is easier when the data are aggregated in a small number of locations that are connected and follow compatible standards for data and metadata storage, i.e., interoperable. As outlined in the use cases described below, the international imaging community needs to address several limiting factors to promote global image data sharing:
- Engage with national and regional funders about the value added by resources that enable image data storage and sharing
- Educate on the importance of quality-managed and harmonized data and rewarding those engaged in the laborious process of producing and curating the data
- Address the technical challenges of moving extremely large datasets
- Find common legal and ethical solutions for sharing medical image datasets that must remain in their country of origin

Balancing these constraints in the long term is possible through a federated model whereby a network of data resources dedicated to particular use cases or domains are stored and made available through a small number of centralized repositories that provide both direct data hosting, as well as indexing and search capabilities. Enabling this model requires development of shared core metadata models for interoperability based on international guidelines such as REMBI (Sarkans et al., 2021), extendable for domain specific applications. Making effective use of the data also requires presentation in standardized formats, particularly those designed for large scale cloud-ready consumption such as OME-Zarr (Moore et al., 2021).

Together, these approaches would allow petabytes to exabytes of immensely valuable scientific data, representing enormous funder investments, to be maximally useful for the global community. Allowing these data to be easily located, accessed through consistent mechanisms in standardised formats and coupled with rich metadata unlocks valuable large-scale applications. These include training universal image classifiers to support automated quality control pipelines for researchers and instrument manufacturers, automated segmentation of common cell structures for use as biomarkers of disease or to better understand mechanisms of action, and prediction of physiological or biomolecular properties of cells to speed up bio-manufacturing, reduce assay variability in labs and reduce consumable use.

However, there are several challenges, which make it more difficult than sharing genomic data or protein structures, for example. One is that individual files may be very large (100 GBs to 10 TBs, and beyond[5]) and require either high-speed internet connections to be uploaded and downloaded or physically transported on external drives. This makes data management and sharing expensive, and it requires storage and maintenance of large datasets over time. A robust technical infrastructure, including large data storage capacity, and powerful computing resources is hence required, as many research institutions still do not have the necessary infrastructure to support image data sharing. Another key challenge is that careful management and organization is needed to ensure that the image data becomes accessible and ideally reusable by other researchers. While medical images typically are easily shared since most are acquired in DICOM format, biological imaging data sharing is still limited because of the lack of standardisation in image data formats and metadata. This makes it difficult for researchers to compare and analyse data from different sources. Ensuring that the data are properly labelled, annotated, and stored in a standardised format can require significant time and resources (Swedlow et al., 2021), which is an essential, yet often underappreciated process. In addition, proper curation and harmonization of imaging data and associated metadata are needed to ensure future merging of data sets for analyses at scale. Finally, depending on the context intellectual property issues, questions of ownership and data privacy concerns may arise. Because of this, researchers may be reluctant to share their image data, and in case of medical imaging data, they must take steps to ensure that the data are de-identified and cannot be traced back to individual patients. Finally, there may be cultural barriers, such as a lack of trust or reluctance to share image data due to concerns about competition or ownership.

Addressing these challenges requires a concerted effort and endurance from researchers, institutions, funding agencies, publishers, and industry to establish internationally recognized best practices for sharing image data in the life sciences. It will also require investments in technical infrastructure and resources, and a shift in the cultural attitudes towards data sharing and collaboration in the scientific community.

In this **White Paper**, we take a first step at presenting some of the most common use cases as well as critical/emerging use cases of (including use of artificial intelligence for) biological and medical image data, which would benefit tremendously from better frameworks for sharing (including technical, resourcing, legal, and ethical aspects). In the second half of this paper, we paint an 'ideal world scenario' for how global image data sharing could work and benefit all life sciences and beyond. As this is still a long way off, we conclude by suggesting several concrete measures directed towards our institutions, existing imaging communities and data initiatives, and national funders as well as publishers. Our vision is that within the next ten years most researchers in this world will be in a position to make datasets openly available as well as access and use quality image data of interest for their research and benefit.

---

[5] https://h01-release.storage.googleapis.com/landing.html

**Table 1: Existing public image data repositories for open data sharing**

| Type | Name | Link |
| --- | --- | --- |
| Public image data repositories | BioImage Archive | https://www.ebi.ac.uk/bioimage-archive/ (Hartley et al., 2022) |
| Public image data repositories | Image Data Resource | https://idr.openmicroscopy.org/ (Williams et al., 2017) |
| Public image data repositories | SSBD, a platform for Systems Science of Biological Dynamics | https://ssbd.riken.jp (Tohsato et al., 2016) |
| Public image data repositories | Medical Imaging and Data Resource Commons, MIDRC | https://midrc.org and https://data.midrc.org/ |
| Public image data repositories and for public download | EMPIAR, the Electron Microscopy Public Image Archive | https://www.ebi.ac.uk/empiar/ (Iudin et al., 2022) |
| For public download | The Cancer Image Archive | https://www.cancerimagingarchive.net/ |
| Cloud-based secure access to cancer imaging data | Imaging Data Commons (IDC) | https://datacommons.cancer.gov/repository/imaging-data-commons |

**Table 2: Relevant publications, recommendations, and guidelines on image data management and sharing (non-exhaustive).**

| Topic | Title | Link |
| --- | --- | --- |
| Public image data repositories | A call for public archives for biological image data (Ellenberg et al., 2018) | https://www.nature.com/articles/s41592-018-0195-8.pdf |
| Public image data repositories; standards for image data formats | A global view of standards for open image data formats and repositories (Swedlow et al., 2021) | https://doi.org/10.1038/s41592-021-01113-7 |
| Data infrastructure (local) | Position statement: Biologists need modern data infrastructure on campus (Andreev et al., 2021) | https://arxiv.org/abs/2108.07631 |
| FAIR image data | Building a FAIR image data ecosystem for microscopy (Kemmer et al., 2023) | https://zenodo.org/record/7788899#.ZE-gJ8FByw4 |
| Image data file formats | OME-NGFF: a next-generation file format for expanding | https://www.nature.com/articles/s41592-021-01326-w |

| | bioimaging data-access strategies (Moore et al., 2021) | |
|---|---|---|
| Metadata guidelines | REMBI (Sarkans et al., 2021) | https://www.nature.com/articles/s41592-021-01166-8.pdf |
| Metadata guidelines | Towards community-driven metadata standards for light microscopy: Tiered specifications extending the OME model (Hammer et al., 2021) | https://www.nature.com/articles/s41592-021-01327-9 |
| Image data formatting and annotation | Community-developed checklists for publishing images and image analysis (Schmied et al., 2023) | https://arxiv.org/abs/2302.07005 |
| Image data sharing and workflows | The new era of quantitative cell imaging—challenges and opportunities (Bagheri et al., 2022) | https://www.sciencedirect.com/science/article/pii/S1097276521010868 |
| Artificial intelligence | BioImage Model Zoo: A Community-Driven Resource for Accessible Deep Learning in BioImage Analysis (Ouyang et al., 2022) | https://www.biorxiv.org/content/10.1101/2022.06.07.495102v1 |

**Table 3: Relevant international communities and initiatives in the field (non-exhaustive).**

| Type | Name | Link |
|---|---|---|
| International community | Global BioImaging | https://globalbioimaging.org |
| International community | BioImaging North America | https://www.bioimagingnorthamerica.org/ |
| International community | Quarep LIMI | https://quarep.org/ https://quarep.org/working-groups/wg-7-metadata/ |
| International community | AI4Life: AI models and methods for the life sciences (image data) | https://ai4life.eurobioimaging.eu/ |
| International community | NEUBIAS - Network of European BioImage Analysts/SoBIAS - Society for Bioimage Analysis | https://eubias.org/NEUBIAS/ |
| International community | vEM: Volume Electron Microscopy | https://www.volumeem.org/#/ |
| International community | ABRF: Association of Biomolecular Resource Facilities | https://www.abrf.org |

| International imaging infrastructure (open access) | Euro-BioImaging ERIC | www.eurobioimaging.eu |
|---|---|---|
| National imaging data initiative | NFDI4BIOIMAGE (Germany) | https://nfdi4bioimage.de/en/start/ |
| National imaging infrastructure (open access) | ABiS: Advanced Bioimaging Support (Japan) | https://www.nibb.ac.jp/abis/ |
| National imaging infrastructure (open access) | Microscopy Australia | https://micro.org.au/ |
| National imaging infrastructure (open access) | National Imaging Facility Australia | https://anif.org.au/ |
| European imaging data initiative | EUCAIM: EUropean Federation for CAncer IMages | https://www.eibir.org/projects/eucaim/ |
| National resource & community (NIBIB[6]-funded) | Medical Imaging and Data Resource Center | https://www.midrc.org/ and https://data.midrc.org/ |
| European FAIR data and service infrastructure | European Open Science Cloud | https://www.eosc.eu/ |
| German National Scientific Data Infrastructure | Multi Disciplinary (Data Science, BioImage, etc. etc.) | https://www.nfdi.de/ |
| Association of Biomolecular Resource Facilities | Membership association | https://www.abrf.org/ |

## 3. Use cases representing different image data types and their challenges and status for sharing

Here, we present different use cases of image data and their potential and challenges in the context of image data sharing. We deliberately choose to present different data types spanning the broad scope from light microscopy and electron to medical imaging, we and conclude with a use case on artificial intelligence being applied to image data. We are convinced that sharing of FAIR image data from all these domains will significantly increase interoperability among the different research domains in the life sciences. Most importantly, it will allow a comprehensive view of biological processes at different levels of organization, from molecular to cellular to whole organism and thereby will advance both basic biological as well as applied health research. The integration of light microscopy and medical imaging can provide a better understanding of disease mechanisms, particularly in the context of complex diseases that involve multiple organ systems and cellular processes. Preclinical and clinical imaging techniques, such as magnetic resonance imaging (MRI), provide valuable data on anatomical structures and physiological functions. Integrating this data with light microscopy images can provide a more accurate and detailed understanding of the underlying biology. This will trigger advancements in drug discovery and development and help to identify potential therapeutic targets, as well as to evaluate drug efficacy and safety. In summary, if we foster image data sharing in the domains of each of the following use cases (and beyond), then we will enable not only promote open science but also new discoveries by being able to link different types of image data sets across the entire spectrum of life.

---
[6] https://www.nibib.nih.gov/

**Use case 1: Light microscopy**

The scale and complexity of light microscopy data area is increasing exponentially as applications become more advanced and automation hardware (motorized stages, objectives, filters, etc.) becomes more affordable and more commoditized. With this increase, so does the challenges of sharing these data. Where images are very large (e.g., light sheet microscopy, intra-vital imaging, hyperspectral imaging, etc.) or datasets have expanded (high content screening and whole slide pathology) on-going discussions as to what the minimal "raw" data storage requirements should be and what "raw" data even means are helping shape the storage landscape for the future. For the majority of the community, the cost of storing data in a shareable, open, and standardized format generally outweighs the cost needed to re-perform the experiment, if it is even possible, e.g., rare patient samples. The richness of these image datasets make them not only valuable for reproducibility, but also ripe for auxiliary analysis, testing other hypotheses, and training AI/ML algorithms. But without clear and easy to use portals that are free or heavily subsidized, individual labs bearing the storage costs and complexities of making data secure and FAIR often see this cost-benefit quite differently from the scientific community at large.

Several repositories support various mixes of data types, metadata requirements, and cost models (Table 1). However, small labs or those that do not have appropriate domain expertise, face large hurdles in using these portals, and often funding is a barrier for long term archiving/sharing of data at the scale that can now be generated by an automated microscope in any life science lab. For those repositories that do not require ongoing user-supported storage cost, alternative funding models must be identified for the long-term health of their resources, likely with help from national or international research organizations and science agencies.

Excellent work is being done to create standards for image metadata and organization schema by groups such as the Open Microscopy Environment (OME[7]), QUAREP-LiMi[8], BINA[9] and others (*see Table 2*); create open file format specifications for data files that range in size from megabytes to petabytes (OME-tiff/OME-Zarr, N5[10]) that are optimal for a variety of storage backends including cloud, parallel file systems, and network attached storage; and consolidate the global community around a common set of standards. Currently, these efforts are supported in a grant/cyclic funding cycle that does not lend itself well to stable and continuous development. New/updated funding models/approaches are needed to keep these critical projects healthy, so that the entire community can stand on a strong foundation.

The metadata standardization effort is challenging in light microscopy because metadata standards must be detailed, yet flexible enough to support vastly different experimental types. For example, the variety of light microscopy modalities, from transmitted light, epifluorescent, confocal, multi-photon and light sheet imaging to spectral methods like Raman and Mass-Spectrometry imaging, require complex equipment, settings, and reagents (often custom created in a lab) to fully capture sufficient information to interpret and reproduce the measurement. This results in the need to capture a large number of parameters, risking researcher exhaustion and highlights the need for heavy investment in simplified tooling for metadata capture and sharing.

---

[7] https://www.openmicroscopy.org/
[8] https://quarep.org/
[9] https://www.bioimagingnorthamerica.org/
[10] https://github.com/saalfeldlab/n5

Fortunately, the research community is beginning to define aspects of "experimental design capture". For example, the Research Resource Identification system (RRID)[11] allows researchers to pull persistent unique identifiers (PUIs) for organisms, biological samples, reagents, and even tools, enabling identification of what was done specifically in a particular experiment. FPbase (Lambert, 2019) has a similar setup for microscopes and optical configurations. However, these tools are still beyond the scope of most researchers to be able to "just use" and thus heavy investment in training and education is needed for these tools (much less new tools to address other gaps).

Directing researchers on which aspects of experimental design *must* be captured has been approached by the REMBI, which attempts to enumerate all relevant fields in bioimaging and bioimage analysis metadata, and complements other valuable ontologies including the 4DN-BINA-OME standard (Hammer et al., 2021), Biological imaging methods ontology (fbbi[12]) NCBI taxonomy[13], the EDAM-Bioimaging ontology (Matúš Kalaš et al., 2020). These efforts in conjunction are starting points around which the community can converge.

The next challenge is how to collect this information. Researchers are busy, typically juggling many competing priorities and demands on their time; therefore, systems that minimize the work needed to capture metadata are critical. A number of tools either have been or are being developed for the capture of microscope hardware details; they include - Micro-Meta App (Rigano et al., 2021), MicCheck (Montero Llopis et al., 2021), FPbase, MethodsJ2 (Ryan et al., 2021), MDEMic (Kunis et al., 2021). A major need is tools to link experimental and image metadata, and to easily capture experimental metadata in a structured and reproducible format. For microscopy metadata capture, an ideal tool would have the ability to a) collect data in structures proposed by international guidelines such as REMBI and use persistent identifiers such as RRIDs, b) easily generate subsets and templates for ease of reuse, and c) easily import and export data. Additional features might include integration with electronic lab notebook (ELN) tools, integration with tools like barcode scanners to allow the inclusion of physical reagents, integration with protocol repositories, integration with figure creation tools, and direct export to systems designed to store (repositories) and/or report on (journals) the data produced.

Finally, we need to ensure that researchers will use such a system. The benefits to data generating labs include improved internal quality control, visibility, and the cross-referencing of one's work by generating reusable data, as well as data access for educational institutions. Training will be required to learn how to use these systems and any capital costs involved. Therefore, strong support and commitment from funding agencies will be required to modernize and implement the "sample, to image, to FAIR data" pipeline.

**Use case 2: Volume electron microscopy**

Volume electron microscopy (volume EM or vEM) describes a set of high-resolution imaging techniques used in biomedical research to reveal the 3D ultrastructure of cells, tissues, and small model organisms at nanometer resolution (Collinson et al., 2023). Typically, heavy metal-stained and resin-embedded specimens are sectioned, imaged, and computationally reconstructed to generate information-rich 3D image volumes that capture the ultrastructure of large fields of view.

---

[11] https://www.rrids.org/
[12] https://www.ebi.ac.uk/ols4/ontologies/fbbi
[13] https://www.ncbi.nlm.nih.gov/taxonomy

In the field of connectomics, neuronal "wiring diagrams" derived from up to petabyte-sized image volumes have profoundly advanced our understanding of the brain. vEM has similarly transformed cell and developmental biology in health and disease in various experimental systems. In 2023, Nature cited vEM as a top technology to watch[14].

In contrast to imaging technologies where recorded signals originate primarily from labelled targets (e.g., fluorescence microscopy), vEM non-specifically captures all resolvable heavy metal-stained features in the volume – membranes, cytoskeletal elements, chromatin structures and large protein complexes. Thus, vEM datasets intrinsically lend themselves to sharing and re-use, since any publication accompanying such data interrogates but a fraction of the information acquired. vEM instrumentation is complex and expensive to run, with imaging experiments lasting days, weeks or even years, so re-use of these data increases return-on-investment. However, vEM comprises multiple related-but-distinct imaging modalities; vEM experiments are often combined with each other and with fluorescence and X-ray microscopy, and they typically include image registration, segmentation and model generation steps. Meaningful and accurate observations require that all these large and disparate datasets as well as related raster- and vector-based files be linked in a reliable and spatially coherent manner. This calls for an improvement on traditional file types, simplistic data sharing models, and visualization methods. Furthermore, vEM datasets range from <1GB to >1PB, so current strategies of storing, sharing and interacting with image data break down.

Without specific resources, support, and clear guidelines and targets, a mandate of data sharing is not feasible. Multimodal vEM data can be rendered meaningless unless it is correctly described at multiple levels beyond just the image file, e.g., biosample, sample, specimen, image layers (label maps, correlated images etc. At the same time, the most advanced and rigorous data sharing models will flounder unless they are accompanied by accessible resources and user interfaces. A recent survey of the vEM community showed that a majority of members are comfortable with no more than point-and-click or text entries/ commands; such findings can help calibrate expectations and shape strategic interventions. It appears that, at least for vEM, unless data sharing requirements are accompanied by an "easy button", resource and time-strapped users will be disincentivized or simply unable to share their valuable data.

The current status of data sharing within the vEM community can be roughly split into two camps: on the one hand, large, well-funded consortia (especially within the connectomics community) have done a remarkable job of uploading gigantic datasets[15], along with intuitive browser-based user interfaces and visualization software[16] - although it is too early to evaluate the true extent of re-use of these data. On the other hand, sharing by smaller groups is improving[17] but still patchy, with many investigators uncomfortable with "giving away free data" and/ or unfamiliar with processes for data upload to public archives. These cultural roadblocks are reinforced by a lack of tangible incentives and formal recognition for data sharers, an entrenched hierarchy between data producing "technical" and data consuming "research" positions (and publications), and a lack of vendor support to ease the structured curation and transfer of images.

There has been progress: connectomics researchers have built paradigm-shifting hardware, file format, and software advances to share massive datasets with interactive tools. Practitioners of "cell biology vEM" are also building plugins and tools that allow for data streaming and easier sharing (Pape et al., 2023). There are several repositories for vEM data, with EMPIAR (hosted at EMBL-EBI) emerging as a primary resource for vEM data that accompany manuscripts. Over a series of meetings, workshops, and efforts from working groups, the vEM community is coalescing around an understanding of the image data and accompanying metadata to be shared based on the REMBI recommendations. A nimble implementation of REMBI for vEM will lay the groundwork for facile and meaningful data sharing amongst various research groups, and indeed some institutions have begun this process to cohere disparate data streams from imaging core facilities.

---

[14] https://www.nature.com/articles/s41592-023-01861-8
[15] https://h01-release.storage.googleapis.com/landing.html
[16] https://h01-release.storage.googleapis.com/gallery.html
[17] http://nanotomy.org/ AND https://openorganelle.janelia.org/datasets

**Use case 3: Medical imaging**

Medical imaging enables visualization of human anatomy and function, usually in a non-invasive manner. Typically housed in Radiology and Radiation Therapy departments, medical images are critical in early detection, diagnosis, prognosis, risk assessment, assessing response to therapy, guiding therapeutic interventions, and most recently, theranostics. Typical imaging modalities include radiographs, computed tomography (CT), ultrasound, magnetic resonance imaging (MRI), and nuclear medicine (e.g., PET and SPECT). Almost all medical images are obtained or can be readily converted into DICOM format thus enabling effective and efficient data transportation, archival, and machine learning.

Radiographic imaging (often called "x-rays") and CT use ionizing radiation to measure the attenuation of kilovoltage photons as they pass through the body. Common examples include chest radiographs, full-field digital mammography (FFDM), and tomosynthesis. CT acquires images at multiple angles allowing for the generation of cross-sectional images, yielding "slices" through the body. Acquisition parameters of energy, slice thickness, projections and angles, and whether contrast is employed (and what type) are examples of data elements needed for harmonization of scans across patients, institutions, and protocols. Nuclear medicine techniques include positron emission tomography (PET) and single-photon emission computed tomography (SPECT) scans. Such imaging acquisitions involve radioactive-labeled agents, with the emitting photons being detected by energy-specific sensors. Depending on the targeted anatomic region and the specific label, nuclear medicine systems can yield quantitative information on physiology, such as metabolism. Ultrasound uses sound waves (not ionizing radiation) to non-invasively image regions of the body. Ultrasound probes (transducers) are implemented on the skin externally; however, they can also be used internally such as in the vaginal cavity to obtain better quality images. Most recently, they are used during surgery to assess disease extent. Ultrasound can image structure as well as function such as echogenicity, tissue stiffness (elastography), and blood velocity. MRI is a non-ionizing 3D imaging technique that incorporates a magnetic field that aligns protons within the body and a radiofrequency current that stimulates the protons, which when turned off, the protons realign with the magnetic field releasing energy subsequently detected by MRI sensors. MRIs can give information on structure, the biochemical nature of tissue, and information on the local environment. Use of intravenous contrast agents are used with temporal MRI acquisitions to yield dynamic-contrast enhanced MRI to yield images with information on vascular uptake and tumor angiogenesis.

Computer vision and AI of medical images has been studied since the 1960's, with the 1980's bringing in computer-aided detection/diagnosis (CAD) as means to extract and merge information from medical images to aid radiologists in their interpretation (Giger et al., 2013; Sahiner et al., 2019). These efforts led to the first FDA-approved CADe system in 1998 (in mammography) (Freer & Ulissey, 2001) and the first FDA-cleared CADx system to aid in cancer diagnosis in 2017 (in breast MRI) (Jiang et al., 2021; Yanase & Triantaphyllou, 2019). Many developments have accompanied the improvements in compute power, storage, and deep-learning technologies. It is important to note that such developments have been greatly facilitated by having medical images acquired in the DICOM format, a world-wide, industry agreement to standardize the acquisition of medical images (Mustra et al., 2008).

Multiple repositories exist for medical images – varying in terms of level of curation, governance, accessibility, and interoperability. Examples include The Cancer Image Archive (TCIA), the Image Data Commons (IDC), and Medical Imaging and Data Resource Commons (MIDRC) (Table 1). To enable the development of trustworthy AI of medical images, best practices in terms of the collection, data models, harmonization, diversity, annotations, and training/testing protocols are critical (Hosny et al., 2018). These are being addressed by current repositories, some of which include educational information on bias and on metrology for AI investigators. It is necessary to avoid "garbage in, garbage out". Note that data collection typically entails diagnostic quality images, which can vary from low quality to high quality images. It is important that the data are curated and organized via a data model so that the end user/developer understands how realistic the data are and from which cohorts can be selected. Many imaging repositories include images and some metadata (clinical and demographic data). However, to conduct multi-modal AI, interoperability of image-based data commons with other data commons (such as those with her, genomics, etc) are necessary. Often such interoperability can technically be accomplished, however, the varying governances of different data commons can hinder efficient implementation. To increase imaging studies available to developers, a federation of data commons and sources may be necessary, potentially using combined centralized and federated commons. Ultimately, medical imaging AI algorithms need to be appropriately evaluated for their specific clinical question, specific clinical claim, and their intended population to move through regulatory (FDA) to clinical practice. It is important to note that AI-enabled medical devices cannot be used "off-label".

**Use case 4: Artificial Intelligence with scientific image data**

During the past few years, the analysis of scientific image data has become ever more reliant on machine learning (ML) methods, and more generally, on modern methods from artificial intelligence (AI). Practically at the microscope, "smart microscopy" and other hardware control applications use AI/ML modules to detect rare events and/or regions of interest (ROIs) at which an imaging device should alter its mode of operation. Downstream from image acquisition, virtually all kinds of analyses now depend on AI/ML techniques, either via methods that are pure ML approaches (e.g. image denoising or segmentation), or hybrid approaches that combine AI/ML modules with additional computational components (e.g. cell or object tracking). Common to all successful applications of AI is the need for adequate amounts of image data and suitable data annotations for AI model training and validation.

Hence, the successful application of AI methods to data-driven life science applications hinges on the availability and accessibility of suitable image and label data. While some applications can be trained through self-supervision, i.e. on raw image data only, like Noise2Void (Krull et al., 2019), most models require external supervision that can stem from a concurrent imaging modality (Bai et al., 2023; Wagner et al., 2021) or from laborious manual annotation by human experts (Greenwald et al., 2022; Stringer et al., 2021). To give one example, the Segment Anything Model (Kirillov et al., 2023) was trained on 11 billion object masks, which someone had to create, collect, and curate before the final model could be trained. The benefits and drawbacks of the two main strategies: training multiple smaller models with a limited range of application or large, foundational models that can be applied to a much broader range of data modalities are still heavily debated in the community. The benefit of smaller models is much better control of the data domain they can operate on, but smaller models must be created again and again for the changing context of life science projects. In contrast, foundational models promise to be more widely applicable to a large range of biological problems at the cost of much higher data and computational resources requirements in training. Additionally, foundational models can be an excellent starting point for more specialized models via processes like fine-tuning, where the original model gets slightly modified for the sake of performing even better on a more specific task at hand.

Another aspect to this debate is that the heterogeneity of training data is key to improved generalisation performance of AI models. As we have learned from large language models such as the GPT model series (Brown et al., 2020; Radford et al., 2019), training on an unfathomable amount of data can yield generalization performance that drives applications like Chat-GPT (OpenAI, 2023). Still, such paradigm shifting foundational models can only be trained after enough sufficiently well annotated data is collected and made available. Hence, the key to truly changing the landscape of foundational model training for scientific image data processing and analysis is a central resource that collects image data from different sources and all relevant scientific image data modalities and that is annotated not only with metadata regarding the imaging process but also with metadata describing the sample and sample preparation.

Metadata of interest extends beyond the aspects mentioned above. Once analysis methods and/or AI models are established, metadata must also include specific image processing information that enables others to reuse a method or model in a valid way (some methods, for example, require an input image to have a number of pixels per dimension that is divisible by a specific number, or they can only be applied to large images when a certain amount of pixel context is given, etc.). Hence, a consistent metadata collection is also important to determine the applicability of pretrained models to new data. Such consistency is crucial to ensure fair and unbiased AI models, especially when dealing with medical image analysis (Drukker et al., 2023).

Once AI models are trained, uniform and standardized (FAIR) access to these pre-trained models for scientific image data analysis needs to be provided. While many collections of trained models have been created for method developers (eg: PyTorch model zoo[18]), other initiatives do also have a user and tool developer perspective in mind (e.g., Hugging Face[19], Bioimage Model Zoo[20]). The Bioimage Model Zoo (Table 1) is a solution built to meet the needs of the life science imaging community, hosting trained models, while also giving credit to all involved parties, linking to training data, teaching material, and consumer software tools that can run the hosted models. Such initiatives will increase the reproducibility and reusability of AI methods and aid their adoption and adaptation by life scientists to their specific analysis needs.

Another important aspect is that models, data, and metadata (see above) need to be programmatically accessible to computer scientists, method and tool developers, and bioimage analysts according to the FAIR principles. This will increase the utility of such data, reduce the duplication of datasets across sites, and therefore save energy, money, and resources. Ideally, we could, as a community, put additional incentives in place such as contributor badges or open calls to the imaging community with prizes or awards to close the gaps in publicly available data. In the long-term, though, we would hope that the reuse of deposited data, data labels, metadata, or trained AI models would lead to resource citations, which would directly fuel the existing performance indicator values used in academia. While we start to see good examples of thorough work combining heterogeneous data with detailed metadata for training ML/DL models (Conrad & Narayan, 2023), recognition metrics will further support a positive change in mindset.

In summary, we believe that the FAIR availability of data and AI models through suitable open data resources, based on community standards for metadata and programmatic access, will greatly increase the rate of new method development. Furthermore, life science experimentalists and imaging scientists will reap the benefits of improved usability and ease of deployment of the new and old AI/ML methods and, through contribution of additional data, will in turn motivate the closing of the remaining gaps in unaddressed tasks and imaging modalities.

---

[18] https://pytorch.org/serve/model_zoo.html
[19] https://huggingface.co/
[20] https://bioimage.io/#/

## 4. Towards global image data sharing

Creating an ecosystem for global image data sharing requires engagement with the relevant stakeholders:

New **funding mechanisms** must be considered for various critical but non-traditional work in this space. "Plumbing" to develop, maintain, and upgrade image data sharing pipelines into seamless data ecosystems must be funded. These include not only software developers who create modules - and importantly, "easy buttons" for users - but also data wranglers and curators who will perform key under-the-hood work, much of which is unlikely to be published. Funders must also identify and support key archives and repositories in various fields, and then enforce image data and metadata formats for users. An international federated system is an appealing model that allows resiliency and easier access/response. For key areas (disease models for example), funding of architectures and benchmark datasets with features such as correlated multimodal images and versioned segmentation will allow data mining and standardization of future experiments in the area. Finally, in the spirit of true democratization, funders must support cloud resources for smaller, resource strapped labs, while training and educating researchers on the benefits of and tools for most efficiently sharing data for maximal impact. Conceptually, for grant proposals, the creation and sharing of meaningful data specifically for re-use must be treated on par with scientific manuscripts - or alternatively, alternative funding streams must be created for such data streams. On a larger scale, funders should consider novel approaches to image data acquisition and sharing, such as national or regional networks. Given the expense of instrumentation and upkeep, advanced imaging lends itself to shared facilities to maximise return-on-investment. To prevent existing and new facilities from devolving into a patchwork of unconnected and variably capable labs, funders must move beyond the traditional "pay for play" core facility and toward imaginative models along the lines of existing national imaging infrastructures as they can be found under the legal entity of Euro-BioImaging ERIC.

**Journals** must require meaningful data sharing and must provide authors with resources to do so. For example, tools to quality control datasets and metadata are critical, so that these resources do not devolve into data dumps. Perhaps similar to Methods and Protocol journals, further support and recognition of "Data Journals" (Kindling & Strecker, 2022), that publish key datasets with some accompanying research to establish utility and quality, will help drive trust and eventual re-use of data This will also pave the way to publish quality data and gain credit, by being able to cite shared data at par with method and protocol papers.

Today's imaging technologies are inseparable from high-end instrumentation and downstream computation. Given the expensive nature of these tools, the relatively slow rate and potentially high volume of data generation, and the intrinsic reusability of imaging data, **vendors**' responsibilities must extend beyond the point of data creation (i.e., writing detector recordings to disk). The image data must be accompanied by metadata that allow thorough descriptions of the images and ideally also experiments leading to them, and the metadata itself should be well structured according to accepted formats. Critically, with recent investments in developing commercial "complete pipelines", users must have the option to easily exit vendor-created silos and export these data and metadata into open formats.

All these ideas gain potency only when there is active adoption at the user and **community** level. The awkward truth is that there are ingrained habits about sharing data; on top of a general reluctance to openly share data that may be re-mined by others, there is also an unfortunate tendency to "check the boxes" during publication and upload a poorly annotated dataset, which is of little use. Some of these behaviours are understandable given the scarcity and modes of funding - too much emphasis on new data generation, and not enough on data sharing and data re-use. That said, it is undeniable that the community must get over these qualms and share data meaningfully - anecdotal evidence suggests that such data actually engenders collaborations and new discoveries. The community can also play a critical role in enabling useful data sharing by coming together to agree on metadata standards. Implementations of recommendations that allow easy metadata field population and simultaneously enable project tracking may incentivize uptake of such tools. Luckily, many international imaging initiatives such as Global BioImaging (here: International Working Group on Image Data), QUAREP-LiMi, BINA and others (*see Table 2*), have been launched in the recent past, which in close collaboration are positioned to take on this critical role and speak on behalf of their communities.

Finally, a word about **artificial intelligence**, which is disrupting a wide variety of scientific and non-scientific fields at an astonishing rate. It is clear that in the near future, much of the querying of large image data will be done by AI, so the design of data sharing platforms and schemas must have this in mind. Creating large image datasets that can be mined for AI algorithm/model training is enticing due to the potential to reveal new insights, but without guardrails, these non-moral computational processes and results can have profound and unforeseeable impacts to the imaging community and beyond, most easily intuited with possible AI-based diagnostic or predictive tools in the clinic. We do not delve deep into the ethics of AI, but two rules of thumb could be: 1. For any model trained on publicly available data, the model and full training pipeline must be transparent and public to prevent "black box" predictions and synthesised data (AI model cards[21] or nutritious labels[22]). 2. Results or images that were generated by AI models must be clearly marked as such. Growing efforts in this direction, like elucidated in the Use Case 4, will also need dedicated and sustained support to grow into a global resource.

---

[21] https://arxiv.org/abs/1810.03993
[22] https://arxiv.org/abs/1805.03677

**Towards Global Image Data Sharing:**

**A to-do list for various stakeholders**

- Launch, identify and support key data repositories. An international federated archival system with some software/compute resources is an appealing model that supports resilience and access.
- Support education of researchers on benefits of image data sharing and reuse, FAIR data, as well as training and dissemination of tools to do the same in a facile manner.
- Develop metrics to recognize and incentivize high-quality, structured data generation that allows easy sharing and re-use.
- Provide resources to build useful, accessible, and community recommendation-based implementations of metadata standards to incentivize uptake and enable enforcement.
- Develop new funding mechanisms for critical but non-traditional work in this space e.g., to develop, maintain, and upgrade data sharing pipelines and ecosystems, as well as for software developers, data wranglers, data curators, and FAIR image data stewards.
- Fund data architectures and benchmark datasets in key areas (e.g., disease models).
- Develop and implement data streaming approaches to support navigation of massive data, such as vEM and correlative/ multimodal image datasets.
- Establish guidelines for journals and develop additional specialist "Data Journals" to support re-use and citation of high value datasets, similar to Methods and Protocols journals.
- Develop novel approaches to image data acquisition and sharing, such as co-ordinated national or regional facility networks, to maximise return-on-investment.
- Engage with vendors to mandate that image data be accompanied by well-structured metadata in accepted formats and within data models that thoroughly describe images, with the option to export both image data and metadata into open formats.
- Support development of guardrails for AI in image data acquisition and analysis to clarify data provenance and discourage "black box" solutions, e.g., models and training pipelines must be transparent, and results or images generated by AI models must be clearly marked as such.

**International Working Group Members who contributed to the discussion and writing of the white paper (in alphabetical order)**

| Name | Affiliation |
| --- | --- |
| Peter Bajcsy | National Institute of Standards and Technology, Gaithersburg, MD 20899, USA peter.bajcsy@nist.gov |
| Sreenivas Bhattiprolu | ZEISS Microscopy Customer Center Dublin, CA 94568, USA sreenivas.bhattiprolu@zeiss.com |
| Katy Borner | Intelligent Systems Engineering, Indiana University, Bloomington, IN 47408, USA. katy@iu.edu |
| Beth Cimini | Imaging Platform, Broad Institute, Cambridge, MA USA. bcimini@broadinstitute.org |
| Lucy Collinson | Electron Microscopy Science Technology Platform, Francis Crick Institute, 1 Midland Road, London, NW1 1AT, UK. lucy.collinson@crick.ac.uk |
| Jan Ellenberg | European Molecular Biology Laboratory, Heidelberg, Germany. jan.ellenberg@embl.de |
| Reto Fiolka | Lyda Hill Department of Bioinformatics, UT Southwestern Medical Center, Dallas, Texas, USA. Reto.Fiolka@UTSouthwestern.edu |

| Maryellen Giger | Department of Radiology and Committee on Medical Physics, University of Chicago, Chicago, IL USA<br>m-giger@uchicago.edu |
|---|---|
| Wojtek Goscinski | National Imaging Facility<br>Australia |
| Matthew Hartley | European Molecular Biology Laboratory, European Bioinformatics Institute, Hinxton, Cambridge, UK<br>matthewh@ebi.ac.uk |
| Nathan Hotaling | National Center for Advancing Translational Science, National Institutes of Health<br>Rockville, MD<br>Nathan.Hotaling@nih.gov |
| **Rick Horwitz (contact for working group)** | Allen Institute for Cell Science<br>Seattle, WA<br>rickh@alleninstitute.org |
| Florian Jug | Fondazione Human Technopole, Milan, Italy<br>flotian.jug@fht.org |
| **Antje Keppler (chair of working group)** | Euro-BioImaging ERIC, Bio-Hub, at EMBL, Heidelberg, Germany<br>antje.keppler@eurobioimaging.eu |
| Anna Kreshuk | European Molecular Biology Laboratory, Heidelberg, Germany,<br>anna.kreshuk@embl.de |
| Emma Lundberg | 1. Stanford University, California, US<br>2. SciLifeLab, KTH Royal Institute of Technology, Stockholm, Sweden<br>emmalu@stanford.edu |
| Aastha Mathur | Euro-BioImaging ERIC Bio-Hub, at EMBL, Heidelberg, Germany<br>aastha.mathur@eurobioimaging.eu |
| Kedar Narayan | 1Center for Molecular Microscopy, Center for Cancer Research, National Cancer Institute, National Institutes of Health, Bethesda 20892, Maryland, USA.<br>2Cancer Research Technology Program, Frederick National Laboratory for Cancer Research, Frederick 21702, Maryland, USA.<br>narayank@mail.nih.gov |
| Shuichi Onami | RIKEN Center for Biosystems Dynamics Research, Kobe, Japan<br>sonami@riken.jp |
| Anne Plant | National Institute of Standards and Technology, Gaithersburg, MD 20899, USA<br>Plant, Anne L. Dr.<br>anne.plant@nist.gov |
| Fred Prior | Department of Biomedical Informatics<br>Department of Radiology<br>University of Arkansas for Medical Sciences<br>Little Rock, AR 72205, USA<br>FWPrior@uams.edu |
| Jason Swedlow | Divisions of Computational Biology and Molecular, Cell and Developmental Biology, University of Dundee, UK<br>jrswedlow@dundee.ac.uk |
| Adam Taylor | Sage Bionetworks, Seattle, WA, USA<br>adam.taylor@sagebase.org |

## Acknowledgements

Disclaimer: Commercial products are identified in this document in order to specify the experimental procedure adequately. Such identification is not intended to imply recommendation or endorsement by the National Institute of Standards and Technology, nor is it intended to imply that the products identified are necessarily the best available for the purpose.
Kedar Narayan is funded by Federal funds from the National Cancer Institute, National Institutes of Health, under Contract No. 75N91019D00024. The content of this publication does not necessarily reflect the views or policies of the Department of Health and Human Services, nor does mention of trade names, commercial products, or organizations imply endorsement by the U.S. Government.